# Another thought experiment in special relativity

**Aleksandar Gjurchinovski**

Department of Physics, Faculty of Natural Sciences and Mathematics
Sts. Cyril and Methodius University,
P. O. Box 162, 1000 Skopje, Macedonia

E-mail: agjurcin@iunona.pmf.ukim.edu.mk, agjurcin@yahoo.com

### ABSTRACT

We present a simple thought experiment in which a beam of light of a finite length (a light-pulse) enters a detection device that consists of a wave detector and a light bulb. We examine the experiment from the reference frame where the device is stationary and from the reference frame where it moves at a constant velocity along the axis of the beam. We use basic special relativity to derive the formulas that show how the length of the light beam is transformed between the reference frames. We further show that the uncommon formulas for length transformation of a light beam do not contradict the usual relativistic result that a uniformly moving non-light object will be Lorentz contracted along the direction of its motion.

Consider a chopped beam of light (a light-pulse) of length $l_0$ incident on a stationary wave detector $D$ (see Fig. 1). The detector $D$ is connected to a switching mechanism that controls a light bulb in such a way that when the detector $D$ is disturbed by an electromagnetic energy, the bulb will glow. Otherwise, the bulb is switched-off. The time required for the whole light beam to pass through the point of location of the detector $D$ is

$$\Delta t_0 = \frac{l_0}{c}, \tag{1}$$

which is the time for which the detector $D$ will be disturbed by the light beam, and, thus, the time for which the light bulb will be switched-on. Here, $c$ refers to the speed of the light beam, and equals the speed of light in vacuum.

Let us explore the experiment in Fig. 1 from a reference frame traveling to the left at a constant velocity $v$. Now, the detection device will move to the right at a constant velocity $v$ (see Fig. 2). Let $\Delta t$ be the time interval required for the light beam to pass through the moving detector $D$. This is the time for which the moving light bulb will glow. For this same amount of time $\Delta t$, the detector $D$ will move a distance $v\Delta t$ to the right, and the light beam will traverse the distance $c\Delta t$ in the same direction. The relation

$$c\Delta t = l + v\Delta t \tag{2}$$

becomes obvious from Fig. 2. Here $l$ denotes the length of the light beam with respect to the reference frame where the detection device is moving at a constant velocity $v$ to the right. While writing Eq. (2), we have taken into account the constant light speed postulate, according to which the light beam will preserve its velocity $c$ in both reference frames [1,2].

The time intervals $\Delta t_0$ and $\Delta t$ are related by the time dilation formula

$$\Delta t = \frac{\Delta t_0}{\sqrt{1 - v^2/c^2}}. \tag{3}$$

Elimination of $\Delta t_0$ and $\Delta t$ from Eqs. (1), (2) and (3) yields

$$l = l_0 \sqrt{\frac{1-v/c}{1+v/c}}. \qquad (4)$$

The result states that if a beam of light of a finite length $l_0$ with respect to the reference frame $S_0$, is observed from a reference frame $S$ that travels relative to $S_0$ at a constant velocity $v$ in the direction which is opposite to the direction of motion of the light beam, then the observer in $S$ will conclude that the length of the light beam will be contracted with respect to the length of the light beam in $S_0$ in accordance with Eq. (4). The result seems to contradict the standard Lorentz contraction formula, but we will show that this is not so.

One may notice an interesting similarity between Eq. (4) and the formula for relativistic longitudinal Doppler effect in the case of a plane harmonic electromagnetic wave [3]. Also, one may notice an important point that the term "proper length" makes no sense in the case of a light beam due to the fact that the speed of the light beam in both $S_0$ and $S$ will remain unchanged and equal to $c$.

If we repeat the same derivation in the case when the setup in Fig. 1 is moving to the left at the same constant speed $v$, we will arrive at the conclusion that the length of the light beam will be elongated in accordance to the formula

$$l = l_0 \sqrt{\frac{1+v/c}{1-v/c}}. \qquad (5)$$

Thus, there will exist an asymmetry between the observations of the length of the light beam made by the observer that moves to the left at a constant speed $v$ and the observer that moves to the right at the same constant speed $v$.

The result that the length of the light beam transforms in a way described by Eqs. (4) and (5) does not contradict the usual relativistic result that a uniformly moving non-light object will be Lorentz contracted along the direction of its motion. In order to show that this assertion is correct, let us explore the setup given in Fig. 3. We have a stationary rigid rod of length $L_0$ that has two detection devices placed at both of its ends. The detection device placed at the left end of the rod is slightly modified. It operates in an opposite way to the one at its right end which is the usual detecting device we have been using previously. That is, the light bulb at the left end will be switched-off in the case when the detector $D_1$ is disturbed by an electromagnetic energy, and will be glowing otherwise. Now, consider the light beam of equal length $L_0$ incident on the detecting devices from the left (see Fig. 3). Before the light beam hits the detector $D_1$ at the left end of the rod, the left bulb will glow, and the right bulb will be switched-off (Fig. 3a). During the time of passage of the light beam through the first detector, both light bulbs will be switched- off (Fig. 3b). At the instant when the light beam will hit the detector $D_2$ on the right, the light beam will leave the first detector, and both light bulbs will be instantaneously switched-on (Fig. 3c).

We will now explore the situation in Fig. 3c with respect to the reference frame where the rod is moving at a constant velocity *v* to the right (see Fig. 4). A careful reader may notice that the situation is equivalent to the one in the famous Einstein's train-embankment thought experiment [4] in which the opposite ends of the train are hit by strokes of lightning. If we make a modification in Einstein's experiment by substituting the train with the rod, and the strokes with the detecting devices, we may conclude that the light bulbs in Fig. 4 will not glow instantaneously, but with a certain time delay $\Delta\tau$. That is, the light bulb at the right end of the moving rod will be switched-on a time

$$\Delta\tau = \frac{Lv}{c^2 - v^2} \qquad (6)$$

later than the light bulb at the left [5]. Here, *L* denotes the length of the moving rod. The time interval $\Delta\tau$ is the time that elapsed between the instant when the light beam leaves the left detector (Fig. 4a) and the instant when the light beam hits the right detector (Fig. 4b). During this time interval $\Delta\tau$, the rod has moved the distance $v\Delta\tau$ to the right, and the light beam has traversed the distance $c\Delta\tau$ in the same direction. Hence, from Fig. 4, we have

$$L - d + v\Delta\tau = c\Delta\tau, \qquad (7)$$

where *d* is the length of the light beam. The length *d* of the light beam in the reference frame where the rod is moving at a constant velocity *v* to the right is related to the length $L_0$ of the light beam in the reference frame where the rod is stationary through Eq. (4). Hence,

$$d = L_0 \sqrt{\frac{1 - v/c}{1 + v/c}}. \qquad (8)$$

We substitute Eqs. (6) and (8) into Eq. (7), and solve the resulting equation for $L$ to obtain

$$L = L_0 \sqrt{1 - \frac{v^2}{c^2}}, \tag{9}$$

which states that the moving rod will be Lorentz-contracted along its velocity vector.

Let us repeat the derivation associated with Fig. 3c with respect to the reference frame where the rod is moving at a constant velocity $v$ to the left (see Fig 5). Again, by making an analogy with Einstein's train-embankment, we may conclude that the light bulb at the left end of the moving rod will be switched-on a time

$$\Delta \tau' = \frac{L'v}{c^2 - v^2} \tag{10}$$

later than the light bulb at its right end, where by $L'$ we denote the length of the moving rod in this case. Notice that the time interval $\Delta \tau'$ is the time that elapsed between the instant when the light beam hits the right detector (Fig. 5a) and the instant when the light beam leaves the left detector (Fig. 5b). During this time interval $\Delta \tau'$, the rod has moved the distance $v\Delta \tau'$ to the left, and the light beam has traversed the distance $c\Delta \tau'$ to the right. From Fig. 5, we have

$$d' = c\Delta \tau' + v\Delta \tau' + L'. \tag{11}$$

Taking into account the expression for $\Delta \tau'$ in Eq. (10), and the formula (5) for length transformation of the light beam in this case,

$$d' = L_0 \sqrt{\frac{1 + v/c}{1 - v/c}}, \tag{12}$$

we arrive at the conclusion that the moving rod will obey the contraction formula

$$L' = L_0 \sqrt{1 - \frac{v^2}{c^2}}, \tag{13}$$

which is identical to Eq. (9). The result shows that the formula for length transformation of a light beam is compatible to the usual Lorentz-contraction formula for a uniformly moving non-light object.

When it comes to teaching special relativity to undergraduates, the task is challenging because the subject is hard to comprehend even in its simplest form. A very effective method to motivate and strengthen student's interest in special relativity is to introduce the subject in an old-fashioned way, that is, by using thought experiments [6]. In this paper, we have used the latter method to derive a rather unfamiliar result that the length of a chopped beam of light obeys transformation formulas different from the usual Lorentz-contraction formula for a non-light object [7]. Furthermore, by making an analogy with Einstein's train-embankment thought experiment, we have demonstrated that the result does not contradict Lorentz-contraction formula. In fact, the result can be used to derive the latter formula.

The method of derivation in this paper belongs to the scope of an introductory course in special relativity, and it could serve as a nice classroom exercise that may improve student's understanding of the fundamental concepts of the theory.

**REFERENCES**

[1] C. Møller, *The Theory of Relativity*, 2nd edition (Clarendon Press, Oxford, 1972).

[2] T. A. Moore, *Six Ideas That Shaped Physics*, Unit R: The Laws of Physics Are Frame-Independent (WCB/McGraw-Hill, Boston, 1998).

[3] See, for example, P. A. Tipler and G. Mosca, *Physics for Scientists and Engineers* (W. H. Freeman & Co., 2004), 5th ed., pp.1276-1277.

[4] A. Einstein, *Relativity -The Special and the General theory* (Wings Books, New York, 1961).

[5] This formula follows directly from the two postulates of special relativity. See, for example, the derivation of Eq. (2.3) in A. I. Janis, "Simultaneity and special relativistic kinematics," Am. J. Phys. **51**, 209-213 (1983).

[6] The reader is referred to W. N. Mathews Jr., "Relativistic velocity and acceleration transformations from thought experiments," Am. J. Phys. **73**, 45-51 (2005), for an extensive list of references.

[7] A typical example that an improper usage of length transformation formulas could lead to a paradox is given in a recent paper J. E. Avron, E. Berg, D. Goldsmith and A. Gordon, "Is the number of photons a classical invariant?," Eur. J. Phys. **20**, 153-159 (1999). The authors have considered the problem of Lorentz invariance of the number of photons in a fictitious rectangular box that encloses a portion of a plane-polarized electromagnetic wave traveling along the positive $x$ axis of the $S$ reference frame. The box was considered to be stationary with respect to the reference frame $S$, and aligned parallel to its $x$ axis. Then, they have used the results from classical electrodynamics and the photon theory of light, and the assumption that the rectangular box obeys the usual Lorentz contraction formula if observed from the reference frame $S'$ that travels along the positive $x$ axis at a constant velocity $v$, to encounter a paradox that the number of photons enclosed in the box is not Lorentz invariant. The authors have incorrectly assumed that the box will be stationary in $S$, and will be moving at a constant velocity $v$ in $S'$, and thus have obviously made a mistake by considering the volume of the rectangular box as a volume of a non-light object that encloses the wave. If they have correctly considered the volume of the box as the volume of the portion of the wave, and thus, that the box is moving at velocity $c$ in both reference frames, and have used the formula for length transformation in Eq. (5) instead of the usual Lorentz contraction formula, they would have arrived at the correct conclusion that the number of photons is indeed Lorentz invariant. That the rectangular box should move at the speed of light in vacuum in both reference frames was also pointed out by Labarthe in a subsequent comment [J. Labarthe, "Einstein's answer to 'Is the number of photons a classical invariant?'," Eur. J. Phys. **20**, L37-L38 (1999)], referring to the ideas given in Einstein's 1905 relativity paper [A. Einstein, "Zur Elektrodynamik bewegter Körper," Ann. Phys. (Leipzig) **17**, 891-921 (1905)]. An interesting discussion about this issue can be found in L. N. Hand and J. D. Finch, *Analytical Mechanics*, (Cambridge UP, Cambridge, 1998), pp. 505-512.

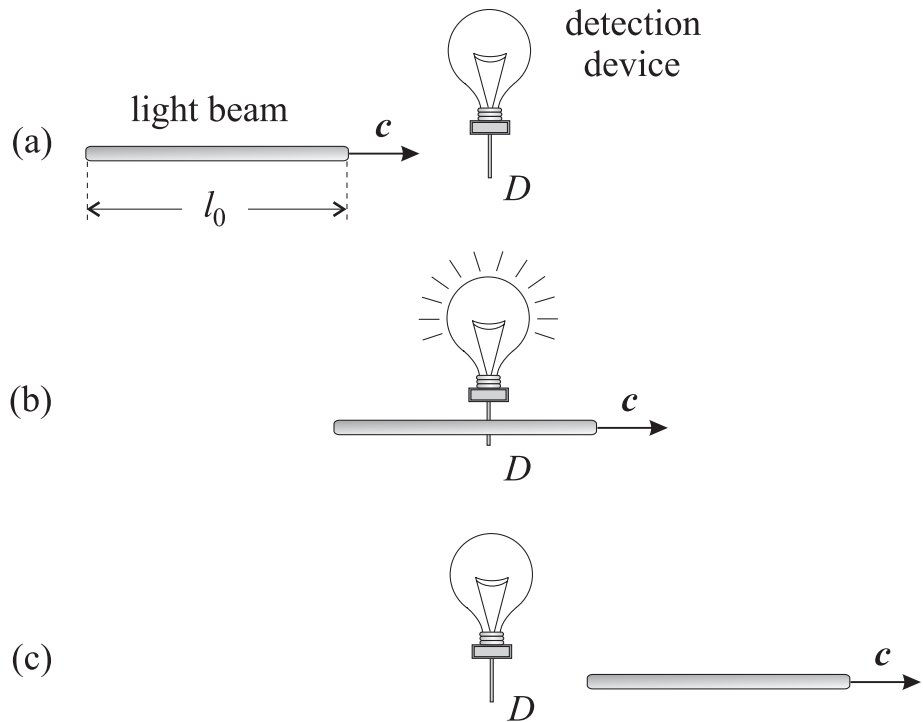

**Fig. 1.** A schematic of the experiment in which a chopped light beam of length $l_0$ enters a detection device that consists of a wave detector $D$ and a light bulb.

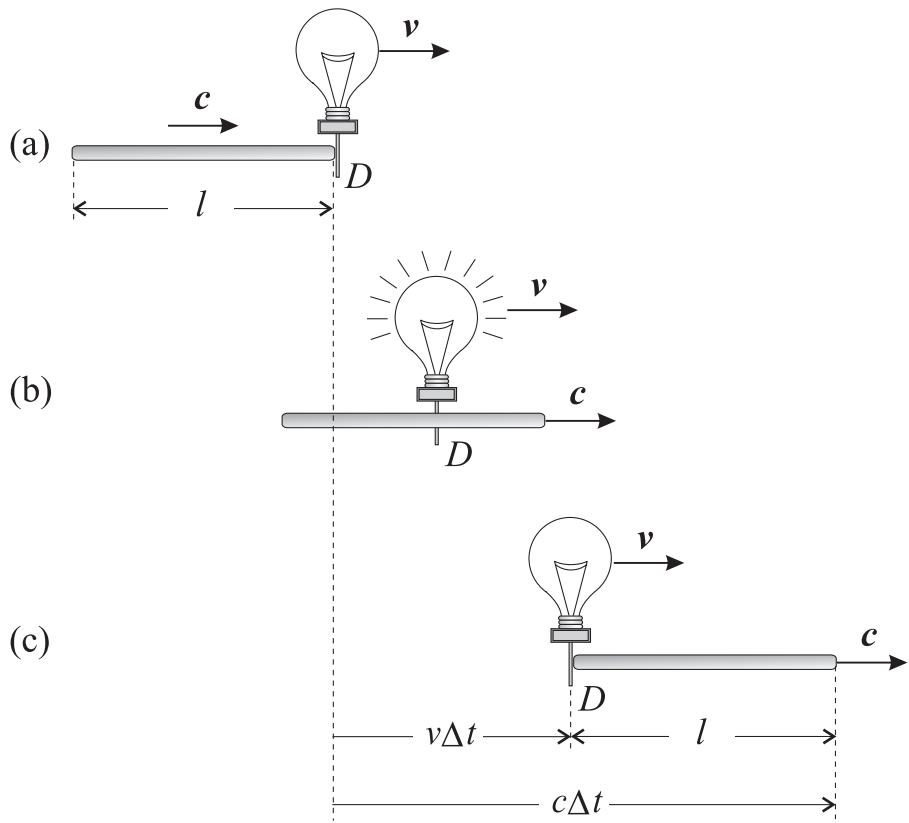

**Fig. 2.** The experiment in Fig. 1 observed from a reference frame traveling to the left at a constant velocity *v*.

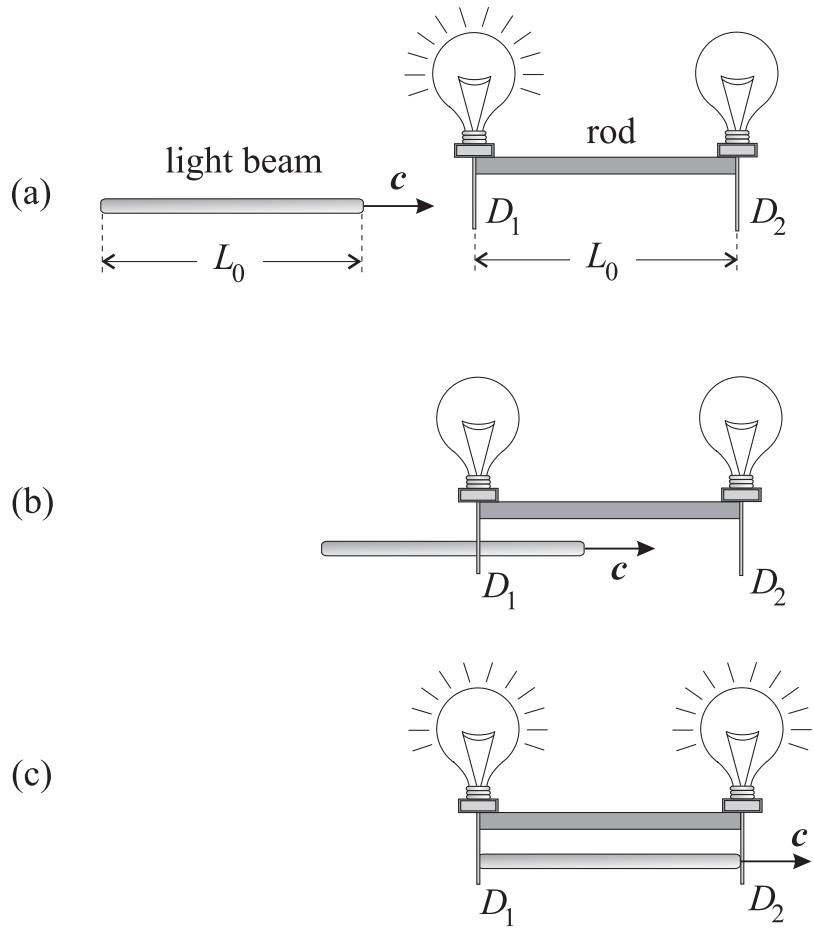

**Fig. 3.** A modified version of the experiment in Fig. 1.

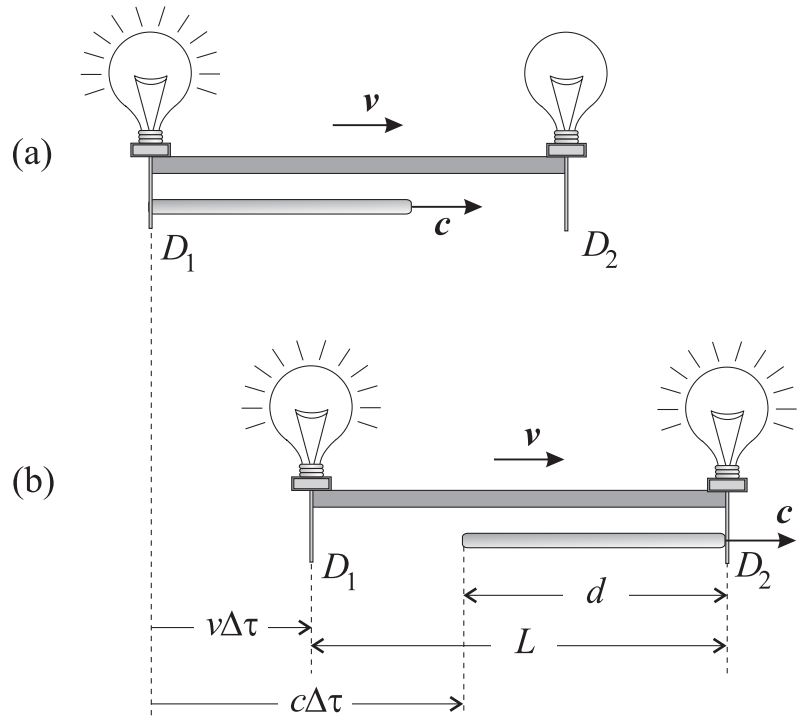

**Fig. 4.** The situation in Fig. 3c observed from a reference frame traveling to the left at a constant velocity *v*. The light beam is shorter than the moving rod.

**Fig. 5.** The situation in Fig. 3c observed from a reference frame traveling to the right at a constant velocity $v$. The light beam is longer than the moving rod.